\documentclass[12pt,preprint]{aastex}

\slugcomment{To appear in ApJ}

\shorttitle{C$_{60}$ in H-rich PNe}
\shortauthors{Garc\'{\i}a-Hern\'andez et al.}

\begin{document}


\title{Formation of fullerenes in H-containing Planetary Nebulae}


\author{D. A. Garc\'{\i}a-Hern\'andez\altaffilmark{1,2}, A.
Manchado\altaffilmark{1,2,3}, P. Garc\'{\i}a-Lario\altaffilmark{4}, L. Stanghellini\altaffilmark{5}, E.
Villaver\altaffilmark{6}, R. A. Shaw\altaffilmark{5}, R.
Szczerba\altaffilmark{7} and J. V. Perea-Calder\'on\altaffilmark{8}}


\altaffiltext{1}{Instituto de Astrof\'{\i}sica de Canarias, C/ Via L\'actea
s/n, 38200 La Laguna, Spain; agarcia@iac.es, amt@iac.es}
\altaffiltext{2}{Departamento de Astrof\'{\i}sica, Universidad de La Laguna (ULL), E-38205 La Laguna, Spain}
\altaffiltext{3}{Consejo Superior de Investigaciones Cient\'{\i}ficas, Spain}
\altaffiltext{4}{Herschel Science Centre. European Space Astronomy Centre,
Research and Scientific Support Department of ESA. Villafranca del Castillo,
P.O. Box 50727. E-28080 Madrid. Spain; Pedro.Garcia-Lario@sciops.esa.int }
\altaffiltext{5}{National Optical Astronomy Observatory, 950 North Cherry
Avenue, Tucson, AZ 85719, USA; shaw@noao.edu, letizia@noao.edu}
\altaffiltext{6}{Departamento de F\'{\i}sica Te\'orica C-XI, Universidad
Aut\'onoma de Madrid, E-28049 Madrid, Spain; eva.villaver@uam.es}
\altaffiltext{7}{N. Copernicus Astronomical Center, Rabia\'nska 8, 87-100
Toru\'n, Poland; szczerba@ncac.torun.pl }
\altaffiltext{8}{European Space Astronomy Centre, INSA S. A.,P.O. Box 50727.
E-28080 Madrid. Spain; Jose.Perea@sciops.esa.int }

\begin{abstract}
Hydrogen depleted environments are considered an essential requirement for the
formation of fullerenes. The recent detection of C$_{60}$ and C$_{70}$
fullerenes in what was interpreted as the hydrogen-poor inner region of a
post-final helium shell flash Planetary Nebula (PN) seemed to confirm this
picture. Here, we present evidence that challenges the current paradigm
regarding fullerene formation, showing that it can take place in circumstellar
environments containing hydrogen. We report the simultaneous detection of
Polycyclic Aromatic Hydrocarbons (PAHs) and fullerenes towards C-rich and
H-containing PNe belonging to environments with very different chemical
histories such as our own Galaxy and the Small Magellanic Cloud. We suggest that
PAHs and fullerenes may be formed by the photochemical processing of
hydrogenated amorphous carbon. These observations suggest that modifications may
be needed to our current understanding of the chemistry of large organic
molecules as well as the chemical processing in space.
\end{abstract}





\section{Introduction}

The current understanding of the fullerene formation is that it is inhibited by
the presence of hydrogen (De Vries et al. 1993; Wang et al. 1995; Cherchneff et
al. 2000; J$\ddot{a}$ger et al. 2009) and as laboratory experiments show, it is
extremely efficient as graphite is vaporized in hydrogen-deficient atmospheres
with helium as buffer gas (Kroto et al. 1985; Kratschmer et al. 1990). As a
consequence,  fullerene molecules in astrophysical domains are expected to be
efficiently formed in hot ($>$3500 K), hydrogen-poor, and C-rich environments
such as Wolf-Rayet (WR) stars (Cherchneff et al. 2000; J$\ddot{a}$ger et al.
2009) or extremely hydrogen-deficient objects such as the R Coronae Borealis
stars (Goeres \& Sedlmayr 1992). In such conditions, fullerenes are thought to
be built from the coalescence of large  monocyclic rings in the gas phase
(Cherchneff et al. 2000) in the absence of Polycyclic Aromatic Hydrocarbon (PAH)
molecules, which are ruled out as possible intermediaries (J$\ddot{a}$ger et al.
2009). Yet to date, no fullerene molecule has been detected in hot WR stars
neither in the most hydrogen-deficient R Coronae Borealis stars
(Garc\'{\i}a-Hern\'andez et al. 2010) despite the expected efficiency of the
formation process in such environments. Moreover, fullerenes are not expected to
be formed in the hydrogen-rich circumstellar envelopes of cool, evolved stars
(e.g., C-rich Asymptotic Giant Branch stars; Herwig 2005) and in the
interstellar medium (De Vries et al. 1993), in spite of the fact that C-rich AGB
stars are the sites of a complex and rich chemistry, with more than dozens of
new complex molecules detected so far (Herbst \& van Dishoeck 2009). In such
cool, dense, and chemically rich conditions the acetylene (C$_{2}$H$_{2}$) and
its radical derivatives are believed to be the precursors of more complex
C-based molecules such as PAHs (Cherchneff \& Cau 1999).

Slow, massive dust-driven mass-loss at the end of the Asymptotic Giant Branch
(AGB), leads  to the formation of optically thick circumstellar envelopes. The
surface chemistry of the  AGB star (Carbon- versus Oxygen-rich based) is
primarily a direct reflection of the  stellar initial mass (which determines the
number of dredge-up processes) and evolutionary stage (Herwig 2005).  The
Planetary Nebula phase starts with the photoionization of the circumstellar
envelope and  represents the immediate stage after the end of the AGB  phase. 
The chemical mix of the ejecta from AGB dredge-up processes is not expected to
be further modified during the PNe stage, apart from dust processing (Kwok et
al. 2001). During the PNe stage however, fundamental chemical drivers are added
into the system: the presence of a strong and evolving UV radiation field, and
fast, tenuous winds that produce shocks.

The infrared detection of C$_{60}$ and C$_{70}$ fullerenes in the Planetary
Nebula (PN) Tc 1 (Cami et al. 2010) has been reported recently. The authors
state that the inner nebular regions of Tc 1 are carbon-rich, hydrogen-poor and
dusty, and forwarded the hypothesis that Tc 1 underwent a late thermal pulse
then presumably caused the ejection of this material, which now makes up the
warm, dusty, and hydrogen-poor PN core where fullerenes are abundant. This
interpretation is not supported by the literature. In fact, we note neither the
Planetary Nebula (Milanova \& Kholtygin 2009; K$\ddot{o}$ppen et al. 1991) nor
its compact core (Williams et al. 2008; Williams 2010, private communication)
nor and the central star (e.g., Mendez et al. 1988) of Tc 1 are H-poor. These
evidences make it very unlikely that Tc 1 underwent a final helium-shell flash
(Iben et al. 1983) and that the environment where the fullerenes have been
observed is thus hydrogen-poor. In addition, Tc 1 has a low mass central star
(M$_{core}$=0.54 M$_{\odot}$; Maciel et al. 2008) and a slightly sub-solar
metallicity (Perinotto et al. 1994), being identified as a slowly evolving type
II PNe. 

In this letter, we present four new detections of C$_{60}$ fullerenes together
with PAHs and very small amorphous carbon grains  in three Galactic and one
Small Magellanic Cloud H-containing PNe\footnote{The presence of hydrogen is
demonstrated by the presence of PAHs and/or by the supporting available
literature on these sources.}, challenging  the current picture that the
presence or absence of hydrogen in this type of carbon-rich environment clearly
determines whether the chemical pathways favor the formation of PAH molecules or
fullerenes as large aromatic species (Cami et al. 2010). The detection of
fullerene in SMP SMC 16 reported in this letter is the first such detection in
an extragalactic source.

\section{Infrared spectra of fullerene-detected PNe}

The infrared spectra of the PNe M 1-20, M 1-12, K 3-54 and SMP SMC 16 presented
here (see Fig. 1) were all acquired with Spitzer/IRS under several General
Observer programs. Program 3633 (PI: M. Bobrowsky) observed  a sample of 40 PNe
in the direction of the Galactic Bulge (Perea-Calder\'on et al. 2009). Program
20443 (PI: L. Stanghellini) includes 157 compact Galactic disk PNe (Stanghellini
et al., in preparation). Finally, program 50261 (PI: L. Stanghellini) was
directed toward  41 extragalactic (thus low-metallicity PNe) in the Magellanic
Clouds (Stanghellini et al. 2007; Shaw et al. 2010)\footnote{In total, the
Spitzer/IRS spectra of $\sim$240 PNe  were inspected for the presence of the
strongest features of the C$_{60}$ and C$_{70}$ complex species.}. All programs
have in common the spectral coverage in the $\sim$5$-$38 $\mu$m range, making
use of different combinations of the Short-Low (SL: 5.2$-$14.5 $\mu$m; 64 $<$ R$
<$ 128), Long Low (SL: 14.0$-$38 $\mu$m; 64 $<$ R $<$ 128), Short-High (SH:
9.9$-$19.6 $\mu$m; R$\sim$600) and Long-High (LH: 18.7$-$37.2 $\mu$m;
R$\sim$600) modules depending on the source brightness at mid-infrared
wavelengths, and assuring that a minimum S/N of $\sim$50 is usually reached.
More detailed descriptions of the Spitzer observations and the data reduction
process can be found in the relevant references and will not be repeated here.
For comparison purposes we also analyze the Spitzer spectrum of Tc 1 from
program GO 3633, previously published (Perea-Calder\'on et al. 2009) in which
solid state C$_{60}$ and C$_{70}$ fullerenes have been reported recently (Cami
et al. 2010). Following the definitions of low, intermediate, and high
excitation in PNe from infrared lines (Stanghellini et al. 2007) we determine
that the targets with detected fullerene are all low excitation PNe. All these
PNe show also broad dust emission features centred at $\sim$11.5 and 30 $\mu$m
and generally attributed to SiC and MgS, respectively (Speck et al. 2009; Hony
et al. 2002). The strongest solid state C$_{60}$ features at $\sim$17.3 and 18.9
$\mu$m (Kratschmer et al. 1990) are clearly detected superimposed on the dust
continuum thermal emission (Fig. 1).

In order to obtain the residual spectra, where dust and gas features may be
easily identified, we have subtracted the dust continuum emission by fitting
five order polynomials between 5 and 22 $\mu$m at spectral locations free from
any dust or gas feature.  We find that  4 PNe show strong fullerene C$_{60}$
features at $\sim$7.0, 8.5, 17.3, and 18.9 $\mu$m. This phenomenon is much more
common than anyone thought, and deserves much more attention to understand the
implications. The four PNe containing fullerene have similarly low excitation,
infrared spectral energy distributions and carbon dust properties (Fig. 1), and
it has been shown (Stanghellini et al. 2007) that progenitors of PNe with
similar characteristics typically are carbon-rich in the mid-to-lower end of the
AGB mass sequence ($\sim$1$-$2 M$_{\odot}$). Figure 2 shows that the C$_{60}$
fullerene features are clearly detected in all sources.  However, the strongest
and isolated C$_{70}$ features at $\sim$12.6 and 14.9 $\mu$m  are only
tentatively detected in M 1-20 and M 1-12. We can neither confirm nor exclude
the possibility of the latter C$_{70}$ emission features in K 3-54 and in SMP
SMC 16 given the much lower resolution in the Spitzer spectrum for these
sources. The intriguing result is that all three Galactic PNe also show weak PAH
features (e.g., those centered at $\sim$6.2, 7.7, 8.6, and 11.3 $\mu$m). Note
that three of the fullerene-detected PNe (K 3-54, M 1-12 and SMP SMC 16) are
compact PNe ($<$4") and the spectral results represent the integration over the
whole nebulae. However, M 1-20 is an extended PN (like Tc 1) and PAHs and
fullerenes are observed together in the inner 4" region (the aperture of
Spitzer). This seems to be in contrast with the conclusions of Cami et al.
(2010), which associates the presence of fullerene features in the Tc 1 spectrum
with an hydrogen-poor region of the PN. The detection of PAHs in the three
Galactic PNe is highlighted in Figure 3, where an enlarged plot from 5 to 16
$\mu$m is shown.

\section{Mid-IR C$_{60}$ fullerene features}

Table 1 lists the wavelength position and width of the four C$_{60}$ fullerene
features as measured in the residual spectra. The derived positions and widths
of the four neutral C$_{60}$ features seen in M 1-20, M 1-12, K 3-54, and SMP
SMC 16 compare very well with those seen in Tc 1. Note that the profiles,
positions and widths of the C$_{60}$ bands indicate that C$_{60}$ molecules are
in a neutral state, being likely trapped on dust grains (Cami et al. 2010). The
$\sim$7.0 $\mu$m feature is blended with an [Ar II] line but the relative
strenghts of the other three neutral C$_{60}$ features at $\sim$8.4, 17.3 and
18.9 $\mu$m are similar for all sources. The C$_{60}$ molecule's excitation
temperature from the population of the upper states of the four vibrational
states are found to be about 425, 546, 681, and 326 K in M 1-20, M 1-12, K 3-54
and SMP SMC 16, respectively (see Cami et al. 2010 for more details about the
method applied)\footnote{We note that if the 8.5 $\mu$m C$_{60}$ feature is
contaminated by the PAH 8.6 $\mu$m band in the Galactic PNe M 1-20, M 1-12 and K
3-54, then slightly different temperatures, would be obtained.}. This means that
the temperature of the C$_{60}$ molecules in the three Galactic PNe is higher
than in Tc 1 (332 K) and that these molecules - which coexist with other
carbon-based species like PAHs - are closer to the central star. However, the
excitation temperature of C$_{60}$ in the extragalactic and low-metallicity PNe
SMP SMC 16 is almost identical to that of Tc 1 (see Fig. 4). Indeed, Tc 1 and
SMP SMC 16 are infrared spectroscopic twins. The only difference is the clear
presence of a broad 6$-$9 $\mu$m emission feature in SMP SMC 16 which is not
seen in Tc 1 and attributable to hydrogenated amorphous carbon (HAC), very small
grains (VSG) or PAH clusters (Tielens 2008; Buss et al. 1993; Rapacioli et al.
2005).

SMP SMC 16 offers the unique opportunity of obtaining a reliable estimation of
the C$_{60}$ content in H-rich circumstellar ejecta because the distance to the
Small Magellanic Cloud is known with good accuracy to be 61 kpc (Hilditch 2005),
and because of the availability of a reliable C atomic abundance from UV spectra
(Stanghellini et al. 2009). From the number of C$_{60}$ molecules\footnote{We
obtained a total number of 9 x 10$^{47}$ C$_{60}$ molecules for the distance of
61 kpc (see Fig. 4).}, a mass of $\sim$5.44 x 10$^{-7}$ M$_{\odot}$ of pure
C$_{60}$ is obtained. From the [S II] $\lambda$$\lambda$6717,6731 $\AA$~line fluxes
(Shaw et al. 2006), we derived an electronic density n$_{e}$ of 10$^{4}$
cm$^{-3}$. Combining this value with the observed H$_{\beta}$ flux (Shaw et al.
2006) and the electronic temperature T$_{e}$ of 11,800 K (Leisy \& Dennefeld
2006), a hydrogen mass of 0.09 M$_{\odot}$ is derived. Combining this mass with
the carbon abundance (Stanghellini et al. 2009), a C mass of $\sim$1.72 x
10$^{-4}$ M$_{\odot}$ is obtained. Therefore C$_{60}$ represents $\sim$0.32\% of
the total carbon in SMP SMC 16. Our estimation is consistent with previous
estimations for C$_{60}$$^{+}$ (e.g., Foing \& Ehrenfreund 1994) from optical
observations and a factor of 5 lower than the rather uncertain $\sim$1.5\%
estimate in Tc 1 (Cami et al. 2010).

\section{Discussion}

Our observations demonstrate that PAHs and fullerenes coexist in the
circumstellar ejecta of low-excitation and H-containing PNe in our Galaxy and in
the Small Magellanic Cloud. This observational result has profound implications
on our understanding of the chemistry of large organic molecules and the
possible routes of chemical processing in space, highlighting the question of
how these large molecules are formed. This is a very difficult question. At
present, the most likely explanation for the simultaneous presence of fullerenes
and PAHs in H-containing environments is that they may be formed by the
photochemical processing of hydrogenated amorphous carbon (HAC) (Scott \& Duley
1996; Scott et al. 1997a). The distribution of components in the mass spectra of
products sputtered from HAC is found to show a complex dependence on the fluence
and on the irradiation hystory of the surface of the grains (Scott et al.
1997a), being consistent with the known sensitivity of HAC solids to thermal and
photochemical modification (Duley 1993). The laboratory IR spectra (e.g., the
relative strength of the IR features) of HACs are known to be strongly dependent
on the physical conditions and HAC's chemical composition (Scott \& Duley 1996;
Scott et al. 1997a; Scott et al. 1997b; Grishko et al. 2001). In particular,
laboratory studies (Grishko et al. 2001) show that HACs may explain the
broad amorphous bands at $\sim$21, 26 and 30 $\mu$m - the latter feature
sometimes very broad - observed in C-rich proto-PNe and evolved PNe (Hony et al.
2002; Kwok et al. 1999; Hrivnak et al. 2000). Interestingly, all
fullerene-detected PNe show the very broad 30 $\mu$m feature generally
attributed to MgS (Hony et al. 2002). Thus, the broad 30 $\mu$m feature
observed in the fullerene-detected PNe may be also related to HACs, which
should be a major constituent in the circumstellar envelope.

Observationally, it is well known that the net result of the increasing UV
irradiation from the AGB phase to the PNe stage is the transformation from
aliphatic to aromatic groups (Kwok et al. 2001; Garc\'{\i}a-Lario \&
Perea-Calder\'on 2003). Infrared emission spectra of HAC also show this
progression in response to the thermal heating (Scott et al. 1997b; Duley 2000).
In relatively massive C-rich sources, this process must be very fast and under
more energetic conditions (e.g., a rapidly changing UV radiation field or strong
post-AGB shocks). However, in low-excitation, low-mass C-rich objects, this
process is postponed to the PNe stage (e.g., a slowly evolving UV irradiation
and weak post-AGB shocks) and takes place slowly enough that we can see the 
HAC's decomposition products (both PAHs and fullerenes) being generated and
co-existing all together. It is to be noted here that the C$_{60}$
molecules seem to emit in the solid phase, being likely trapped on dust grains
(Cami et al. 2010). Thus, the formation of fullerenes may be facilitated when
the hydrogens have been removed from the surface of the carbonaceous grains. The
de-hydrogenation of the grains is not the consequence of a H-poor environment,
but of the photochemical processing of HACs. We believe this is the case for
the fullerene-detected PNe presented here, showing HACs and/or PAHs and
fullerenes in their circumstellar envelopes. However, the possible evolutionary
sequence of the HAC's decomposition products seen in the PNe is unclear.
Laboratory experiments (Scott et al. 1997a) show that at low fluence conditions
(e.g., like in our slowly evolving PNe), the dissolution of HACs occurs
sequentially, with small molecules and molecular fragments being sputtered
before heavier molecules and clusters. Future observations of a larger sample of
slowly-evolving PNe with different UV irradiation as well as laboratory
spectroscopy of HAC films under very different physical conditions and chemical
composition will help to solve this puzzle. 

In summary, both PAHs and fullerenes may be formed by the decomposition of HAC
(Scott \& Duley 1996; Scott et al. 1997a). Hydrogen is needed to form HAC
grains, which may be then destroyed by the central star's UV photons and/or by
the post-AGB shocks. The products of destruction of HAC grains are PAHs and
fullerenes in the form of C$_{50}$, C$_{60}$, and C$_{70}$ molecules (Scott et
al. 1997a). The C$_{60}$ molecules may be hardy enough to survive for longer
periods of time. This picture would explain why the C$_{60}$ and C$_{70}$
fullerenes in Tc 1 (Cami et al. 2010) are unaccompanied by HACs and PAHs.
Indeed, the very low-excitation and low-metallicity PN SMP SMC 16 shows a
fullerene dominated spectrum with no signs of PAHs; the same is observed in its
Galactic counterpart Tc 1. In addition, this interpretation is supported by the
recent detection of hotter C$_{60}$ molecules (possibly in the gas-phase) in the
least H-deficient R Coronae Borealis stars DY Cen and V854 Cen
(Garc\'{\i}a-Hern\'andez et al. 2010).\footnote{The recent non-detection of
C$_{60}$ across the R Coronae Borealis stars sample (except for the least
H-deficient stars DY Cen and V854 Cen with H-deficiencies of $\sim$10$-$100
only) is additional evidence that C$_{60}$ is not easily (or at all) formed in
H-poor environments (Garc\'{\i}a-Hern\'andez et al. 2010).} Our detection of
fullerenes  in a variety of PNe that span a large range of progenitor's
characteristics yet are narrowed down to a particular mid-infrared spectral 
type and in the typical presence of PAH features, will be an essential starting
point to understand the photochemical  processing of dust in circumstellar (and
perhaps, interstellar) environments. 



\acknowledgments

D.A.G.H acknowledges N. Kameswara Rao and David L. Lambert for the construction
of the excitation diagrams and very interesting discussions. D.A.G.H. and A.M.
also acknowledges support for this work provided by the Spanish Ministry of
Science and Innovation (MICINN) under a JdC grant and under grant
AYA$-$2007$-$64748. Thanks to James Davies for helping in data analysis. This work
is based on observations  made with the Spitzer Space Telescope, which is
operated by the Jet Propulsion Laboratory,  California Institute of Technology,
under NASA contract 1407. L.S. and R.A.S.  acknowledge support by NASA
through awards for programs GO 20443 and 50261 issued by JPL$/$Caltech.  R.Sz.
acknowledges support from grant N203 511838 from Polish MNiSW.



{\it Facilities:} \facility{Spitzer:IRS}.

\clearpage

\begin{deluxetable}{cccccccccccc}
\tabletypesize{\scriptsize}
\tablecaption{Mid-IR C$_{60}$ features$^{a}$ in PNe.\label{tbl-2}}
\tablewidth{0pt}
\tablehead{
\colhead{Feature} &  \colhead{$\lambda$$_{lab}$$^{b}$} &  \colhead{$\lambda$$_{obs}$} & \colhead{FWHM$_{obs}$} &
\colhead{$\lambda$$_{obs}$} & \colhead{FWHM$_{obs}$} &\colhead{$\lambda$$_{obs}$} & \colhead{FWHM$_{obs}$} &
\colhead{$\lambda$$_{obs}$} & \colhead{FWHM$_{obs}$} & \colhead{$\lambda$$_{obs}$} & \colhead{FWHM$_{obs}$}\\
\colhead{} & \colhead{$\mu$m} & \colhead{$\mu$m} & \colhead{$\mu$m} & \colhead{$\mu$m} & \colhead{$\mu$m} & \colhead{$\mu$m} 
& \colhead{$\mu$m} & \colhead{$\mu$m} & \colhead{$\mu$m} & \colhead{$\mu$m}
}
\startdata
 &   & Tc 1 &  & M 1-20  & & M 1-12 & & K 3-54 & & SMC 16 & \\
\hline
T$_{1u}$(4) &  7.00  &  7.03  &  0.193  &  7.02    &  0.178   &  6.99  &  0.092 &  7.03  &  0.344 & 7.02 & 0.129 \\
T$_{1u}$(3) &  8.45  &  8.51  &  0.240  &  8.58    &  0.343   &  8.58  &  0.237 &  8.50  &  0.686 & 8.56 & 0.309 \\
T$_{1u}$(2) & 17.33  & 17.39  &  0.372  & 17.35    &  0.470   & 17.37  &  0.357 & 17.38  &  0.463 &17.42 & 0.614 \\
T$_{1u}$(1) & 18.94  & 18.89  &  0.394  & 18.96    &  0.412   & 18.94  &  0.348 & 18.86  &  0.484 &18.92 & 0.456 \\
\enddata
\tablenotetext{a}{Positions and widths measured in the residual spectra. Note that the 7.0 $\mu$m C$_{60}$  band is blended with [Ar II] 6.99$\mu$m.}
\tablenotetext{b}{Solid state laboratory data from Kratschmer et al.(1990).}
\end{deluxetable}

\clearpage
 
\begin{figure}
\includegraphics[angle=0,scale=.60]{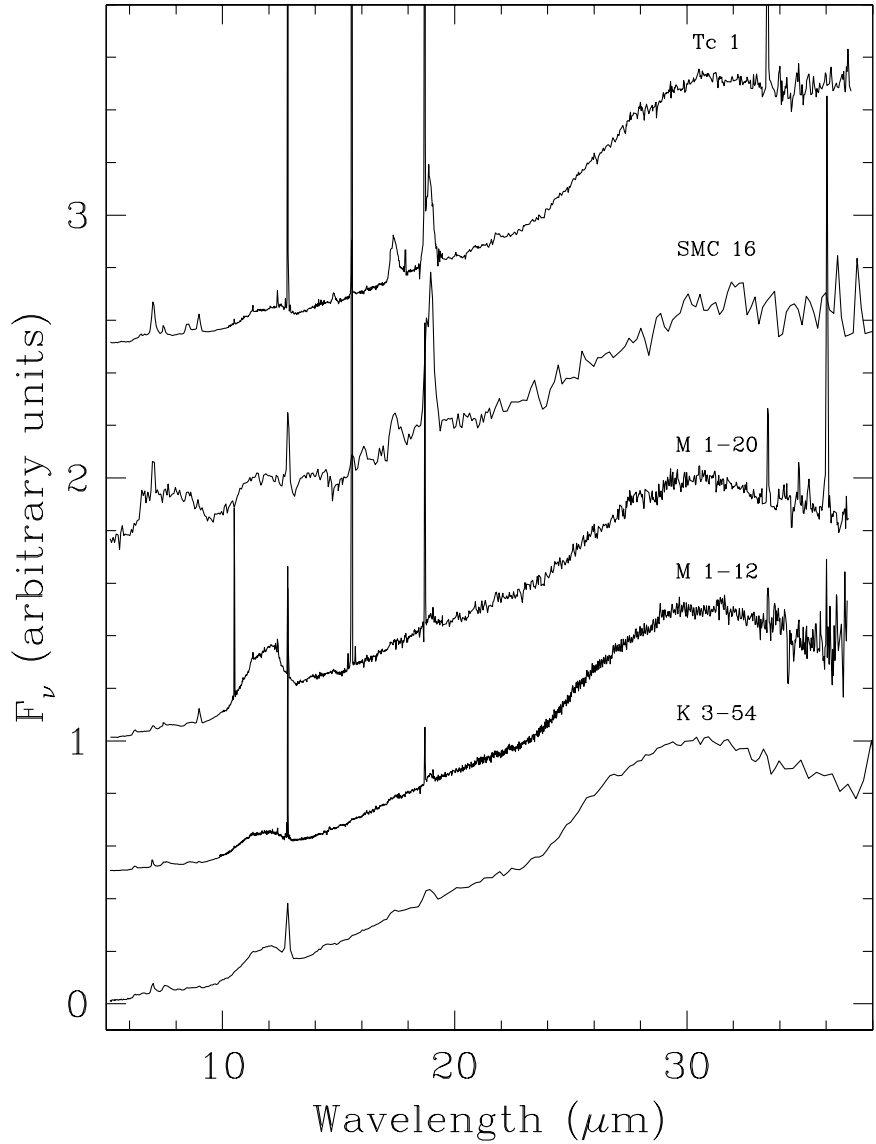}
\caption{Spitzer/IRS spectra in the wavelength $\sim$5$-$38 $\mu$m for the
fullerene-detected PNe Tc 1, SMC SMP 16, M 1-20, M 1-12, and K 3-54. The
spectra are normalized at 30 $\mu$m and displaced for clarity. Note that the
strongest solid state C$_{60}$ features at $\sim$17.3 and 18.9 $\mu$m
(Kratschmer et al. 1990) are clearly detected superimposed on the dust
continuum thermal emission. \label{fig1}}
\end{figure}

\clearpage

\begin{figure}
\includegraphics[angle=0,scale=.60]{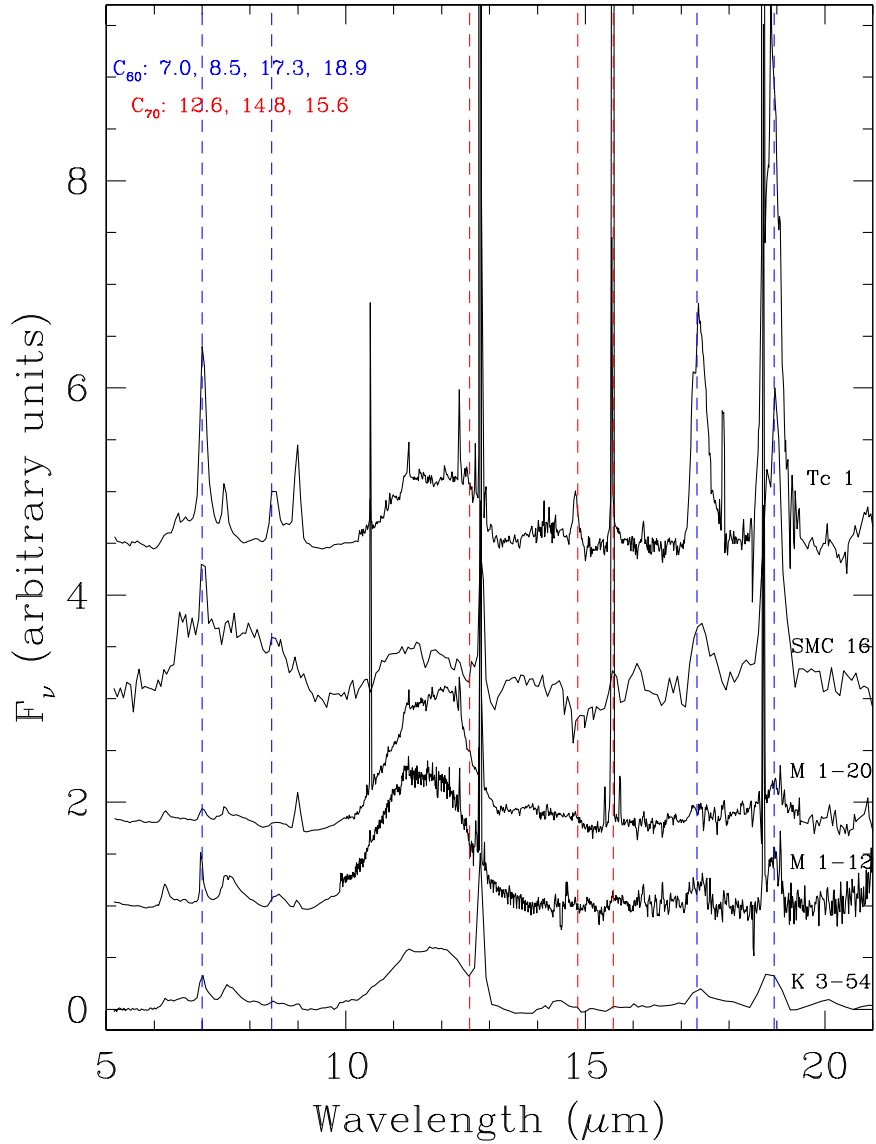}
\caption{Residual spectra in the wavelength range $\sim$5$-$20 $\mu$m for the
PNe Tc 1, SMP SMC 16, M 1 20, M 1-12, and K 3-54. The wavelength positions of
the solid state C$_{60}$ features (Kratschmer et al. 1990) are marked with
blue dashed vertical lines. In addition, the strongest and unblended 
C$_{70}$ features (von Czarnowski \& Meiwes-Broer 1995) are marked with red
dashed vertical lines. \label{fig2}}
\end{figure}

\clearpage

\begin{figure}
\includegraphics[angle=0,scale=.60]{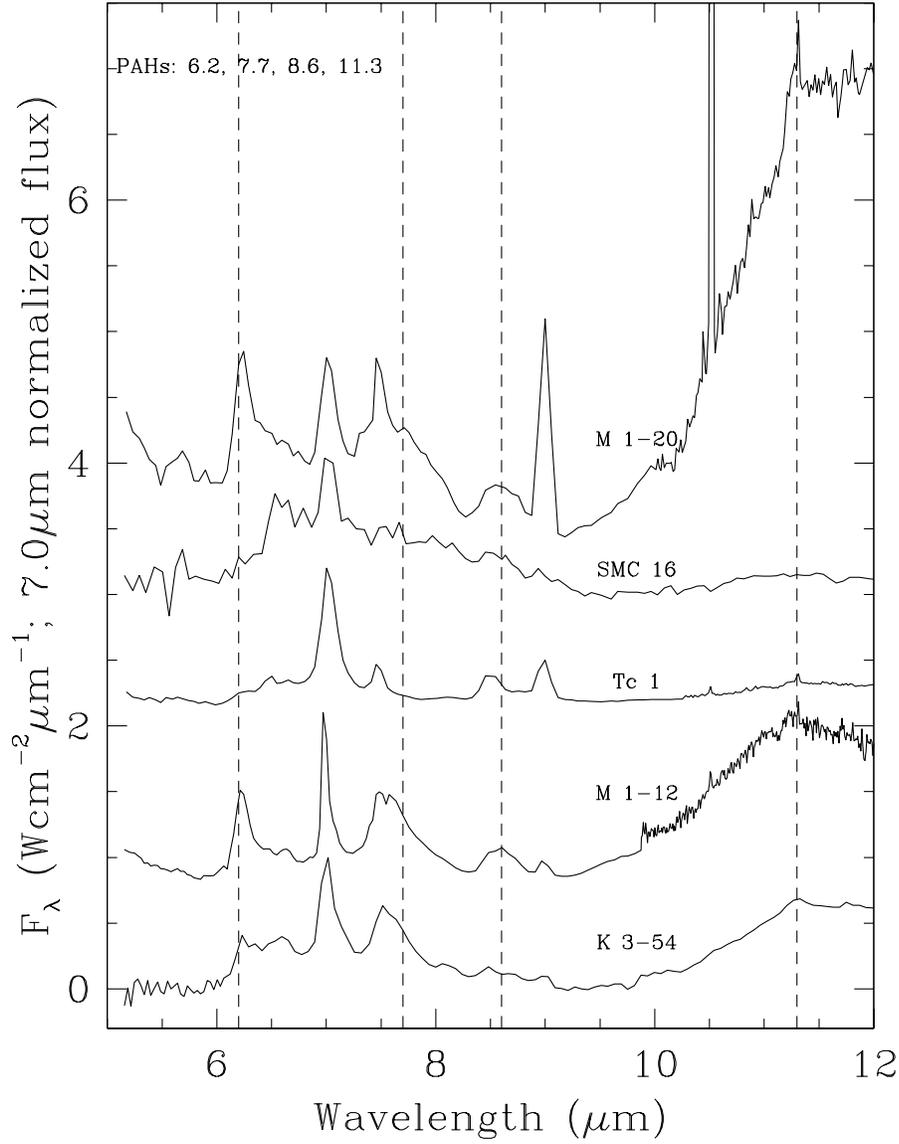}
\caption{Residual spectra in the wavelength range $\sim$5$-$16 $\mu$m for
the PNe Tc 1, SMP SMC 16, M 1-20, M 1-12, and K 3-54. The wavelength
positions of the classical PAH features (6.2, 7.7, 8.6, and 11.3 $\mu$m) are
marked with black dashed vertical lines.\label{fig3}}
\end{figure}

\clearpage

\begin{figure}
\includegraphics[angle=0,scale=.60]{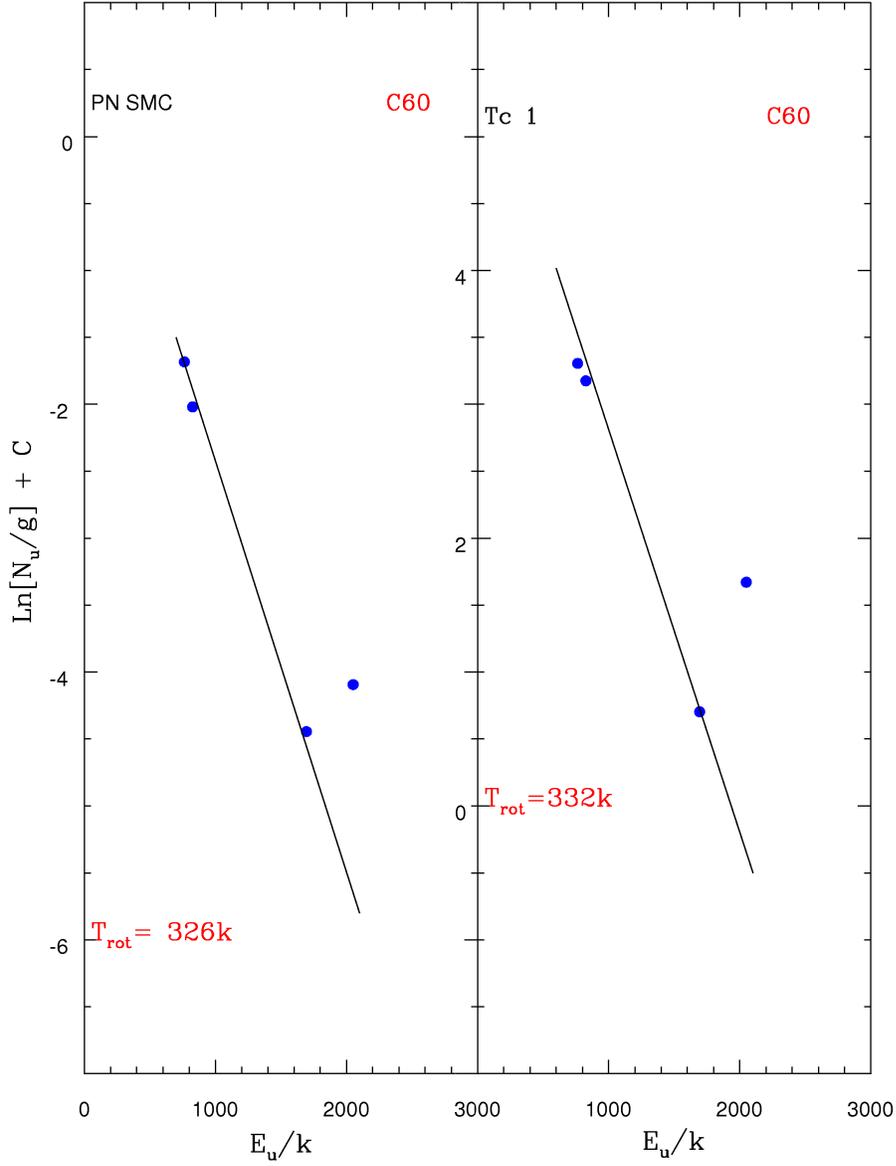}
\caption{Illustrative example of the vibrational excitation temperature
diagrams for the C$_{60}$ bands observed in the PNe SMP SMC 16 (left
panel) and Tc 1 (right panel). \label{fig4}}
\end{figure}

\end{document}